\documentclass[3p,times,review]{elsarticle}

\usepackage{ecrc}
\usepackage{float}
\floatstyle{ruled}
\newfloat{algorithm}{tbp}{loa}
\floatname{algorithm}{Algorithm}

\volume{00}

\firstpage{1}

\journalname{ Digital Signal Processing}

\runauth{}

\jid{procs}

\jnltitlelogo{ Digital Signal Processing}

\CopyrightLine{2011}{Published by Elsevier Ltd.}

\usepackage{amssymb}

\usepackage[figuresright]{rotating}

\begin{document}

\begin{frontmatter}



\dochead{}

\title{Gauss-Newton Filtering incorporating  Levenberg-Marquardt Methods for Radar Tracking}


\author{Roaldje Nadjiasngar}
\address{Roaldje.Nadjiasngar@uct.ac.za}
\author{ Michael Inggs}

\address{Michael.Inggs@uct.ac.za}

\begin{abstract}
This paper shows that the  Levenberg-Marquardt Algorithms (LMA) algorithms can be merged into the Gauss Newton Filters (GNF) to track difficult, non-linear trajectories, without divergence. 
The GNF discusssed in this paper is an iterative filter with memory that was introduced by Norman Morrison \cite{Morrison:1969}. The filter uses back propagation of the predicted state to compute the Jacobian matrix over the filter memory length. The LMA are optimisation techniques widely used for data fitting \cite{ Mquardt}. These  optimisation techniques are iterative and guarantee local convergence. We also show through simulation studies that this filter performance is not affected by the process noise whose knowledge is central to the family of Kalman filters.
\end{abstract}


\end{frontmatter}


\section{Introduction}

This paper shows that the  Levenberg-Marquardt Algorithms (LMA) can be merged into the Gauss Newton Filters (GNF) to track difficult, non-linear trajectories, without  divergence .
\cite{EKFdivergence, blackman,familyKalmanddivergence,divergenceQ1}. In the past, the LMA has been used for initialising tracking filters 
\cite{LMAtrackinitiation,LMAtrackinitiation2,LMAtrackinitiation3}. In this paper we show that the LMA can be merged into the flexible GNF filters to produce a hybrid formulation with very powerful convergence  properties even in highly non-linear input data situations. The hybrid filter is also self initialising.

 The LMA are optimisation techniques widely used for data fitting \cite{ Mquardt}. These  optimisation techniques are iterative and guarantee convergence in a specified region i.e. they do necessarily produce global minima\cite{Ma20071032}. They are also used in most neural networks algorithm \cite{ScienceDLMapplications,ScienceDLMapplications2,ScienceDLMapplications3}.
 
 The Gauss Newton filter (GNF) discussed in this paper was introduced by Morrison\cite{Morrison:1969} to tracking and smoothing at about the same period as the Kalman filter, but it received little attention due to its computational requirements,  problematic for the limited computers of the time. The GNF is iterative and non-recursive, with memory that can be adaptively  controlled. The GNF differs from the Gauss-Newton  optimisation methods discussed in the literature as it provides a different method for computing the Hessian matrix \cite{ScienceDNewtonQuasi}.   This flexibility makes the GNF filter highly suitable for tracking in strongly non-linear situations. 
 
 In this paper we adapt  the GNF to the LMA method (which we call the Morrison LMA Filter) and we state that this filter can be used for radar target tracking without the risk of divergence. We also show through simulation studies that this filter performance is not affected by the process noise whose knowledge is central to the family of Kalman filters. The literature on the use of LMA as a tracking algorithm are rare, possibly due to lack of exposure to Morrison approach in the GNF.
 
The LMA is well known as an aid for track initiation \cite{LMAtrackinitiation,LMAtrackinitiation2,LMAtrackinitiation3}. We make it clear here that the LMA is not applied as an initiation tool in our hybrid filter, but rather as an integral part of the filter. The paper starts in Section \ref{sec:stateSpace} to define a state space model based on nonlinear differential equations.

Section \ref{sec:adaption} is important as it describes the incorporation of the LMA methods into the GNF to produce the Morrison LMA, which converges very robustly . The performance of the new filters is demonstrated in a series of simulations described in Section \ref{sec:simulations}. The paper concludes with a summary and indication of future work.

\section{State space model based on nonlinear differential equations\label{sec:stateSpace}}

Consider the following autonomous, nonlinear differential equation (DE) governing 
the process state:

 \begin{equation}
DX(t)=F(X(t))\end{equation}

in which $F$ is a non linear vector function of the state vector $X$ describing a process, such as the position of a target in space.
We assume the observation scheme of the process is a nonlinear function
of the process state  with expression :

\begin{equation}
Y(t)=G(X(t))+v(t)\end{equation}

where $G$ is a nonlinear function of $X$ and $v(t)$ is a random Gaussian vector. 
The goal is to estimate the process state from the given state nonlinear models. 
For linear DEs,  the state transition matrix  could be easily obtained. This, however, is not the case with a nonlinear DEs. Nevertheless, there is a 
procedure, based on local linearisation, that enables us to get around this obstacle, which we will now present.

\subsection{The method of local linearisation}
The solution of the DE gives rise to infinitely many trajectories that are dependent 
on the initial condition. However there will be one trajectory whose state vector
the filter  will attempt to  identify from the observations. We assume 
that there is a known nominal trajectory with state vector $\bar{X}(t)$  that has the following properties:

 \begin{itemize}
	\item $\bar{X}(t)$ satisfies the same DE as  $X(t)$ 
	\item $\bar{X}(t)$ is close to $X(t)$
\end{itemize}

The above-mentioned properties result in the following expression:\cite{ Mquardt}.

\begin{equation}
X(t)=\bar{X}(t)+\delta X(t) \end{equation}

where $\delta X(t)$ is a vector of time-dependent functions that are small 
in relation to the corresponding elements of either $\bar{X}(t)$ or $X(t)$,
The vector $\delta X(t)$ is called the {\em perturbation vector} and is governed by the 
following DE  (The derivation is shown in Appendix A):

 \begin{equation}
D(\delta X(t))=A(\bar{X}(t))\delta X(t)\end{equation}

where $A(\bar{X}(t))$ is called a sensitivity matrix  defined as follows:

\begin{equation}
A(\bar{X}(t))=\left.\frac{\partial F(X(t))}{\partial(X(t))}\right|_{\bar{X}(t)}.
\end{equation}. 

This equation is therefore a linear DE, with a time varying coefficient and has a 
the following transition equation:

\begin{equation}
\delta X(t+\zeta)=\Phi (t_{n}+\zeta,t_{n},\bar{X})\delta X(t)subsec:local
\end{equation}

in which  $\Phi (t_{n}+\zeta,t_{n},\bar{X})$ is the transition matrix from time 
$t_{n}$ to $t_{n}+\zeta$  (increment $\zeta$). The transition matrix is \cite{ Mquardt}.
governed by the following DE:

\begin{equation}
\frac{\partial}{\partial\zeta}\Phi(t_{n+\zeta},t_{n},\bar{X})=
A(\bar{X}(t_{n}+\zeta))\Phi(t_{n+\zeta},t_{n},\bar{X})
\end{equation}

\begin{equation}
\Phi(t_{n},t_{n},\bar{X})=I 
\end{equation}

The transition matrix is a function of $\bar{X}(t)$  and can be evaluated by 
numerical integration and in order to fill the values of $A(\bar{X}(t_{n}+\zeta))$,
 $\bar{X}(t)$ has to be integrated numerically.
\subsection{The observation perturbation vector }
In this section we will adopt the notation $X_{n}$ and $Y_{n}$ for $X(t_{n})$
 and $Y(t_{n})$ respectively.
We define a simulated noise free observation vector $\bar {Y}_{n}$  as follows:

\begin{equation}
\bar {Y}_{n}=G(\bar {X}_{n}) 
\end{equation}

Subtracting $\bar {Y}_{n}$ from the actual observation $Y_{n}$ gives the  
{\em observation perturbation vector}:

\begin{equation}
\delta Y_{n}=Y_{n}-\bar {Y}_{n}
 \end{equation} 

In Appendix A we show that the observation perturbation vector is related to the 
state perturbation vector as follows:

\begin{equation}
\delta Y_{n}=M(\bar{X}_{n})\delta X_{n}+v_{n} 
\end{equation} 

where $M(\bar{X}_{n})$ is the Jacobean matrix of G, evaluated at $\bar{X}_{n}$. 
The matrix is also called the  {\em observation sensitivity matrix} and is defined as follows: 

\begin{equation}
M(\bar{X}_{n})=\left.\frac{\partial F(X_{n})}{\partial(X_{n})}\right|_{\bar{X}_{n}}
\end{equation}
We now examine the sequence of observations.
\subsection{Sequence of observation}
We assume that $L+1$ observation are obtained with time stamps 
$t_{n}, t_{n-1},...,t_{n-L}$,Theses observations are assembled as follows :

\begin{equation}\label{eq:obs}
\left[\begin{array}{c}
\delta Y_{n}\\
\delta Y_{n-1}\\
.\\
.\\
.\\
\delta Y_{n-L}\end{array}\right]=\left[\begin{array}{c}
M(\bar{X}_{n})\delta X_{n}\\
M(\bar{X}_{n-1})\delta X_{n-1}\\
.\\
.\\
.\\
M(\bar{X}_{n-L})\delta X_{n-L}\end{array}\right]+\left[\begin{array}{c}
v_{n}\\
v_{n-1}\\.
.\\
.\\
.\\
v_{n-L}\end{array}\right]
\end{equation}

Using the relationship: 
\begin{equation}
\delta X_{m}=\Phi (t_{m},t_{n},\bar{X})\delta X_{n}
\end{equation}

then, substituting Equation\ \ref{eq:obs} the observation sensitivity equation can be written as :
\begin{equation}
\mathbf{\delta Y}_{n}=\mathbf{T}_{n}\delta X_{n}+\mathbf{V}_{n}
\end{equation}

in which  $\mathbf{T}_{n}$, the {\em total observation matrix} is defined as follows:

\begin{equation}
\mathbf{T}_{n}=\left[\begin{array}{c}
M(\bar{X}_{n})\\
M(\bar{X}_{n-1})\Phi(t_{n-1},t_{n};\bar{X})\\
.\\
.\\
.\\
M(\bar{X}_{n-L})\Phi(t_{n-L},t_{n};\bar{X})\end{array}\right]
\end{equation}

The vectors  $\mathbf{\delta Y}_{n}$ and $\mathbf{V}_{n}$ are large. 
The cost function we would want to minimize is :
\begin{equation}ef
E(\delta X_{n})=(\mathbf{\delta Y}_{n}-\mathbf{V}_{n})^{T}\mathbf{R}_{n}^{-1}(\mathbf{\delta Y}_{n}-\mathbf{V}_{n})=\delta X_{n}^{T}(\mathbf{T}_{n}^{T}\mathbf{R}_{n}^{-1}\mathbf{T}_{n})\delta X_{n}\label{eq:cost_func}\end{equation}

The solution that minimises the cost function can be obtained  from the minimum variance estimation as follows:
\begin{equation}
\delta\hat{X}_{n}=(\mathbf{T}_{n}^{T}\mathbf{R}_{n}^{-1}\mathbf{T}_{n})^{-1}
\mathbf{T}_{n}^{T}\mathbf{R}_{n}^{-1}\mathbf{\delta Y}_{n}
\label{eq:deltax}\end{equation}
The estimate $\delta\hat{X}_{n}$ has a covariance matrix:

\begin{equation}
S_{n}=(\mathbf{T}_{n}^{T}\mathbf{R}_{n}^{-1}\mathbf{T}_{n})^{-1}
\end{equation}
where $\mathbf{R}_{n}^{-1}$ is a block diagonal weight matrix, also called the {\em least 
squares weight matrix}, but, in fact, if we define $R_{n}$ as the covariance matrix of 
the the error vector $v_{n}$, then $\mathbf{R}_{n}^{-1}$  is expressed as:

\begin{equation}
\mathbf{R}_{n}^{-1}=\left[\begin{array}{cccccc}
R_{n}^{-1} & 0 & . & . & . & 0\\
0 & R_{n-1}^{-1} &  &  &  & .\\
. &  & . &  &  & .\\
. &  &  & . &  & .\\
. &  &  &  & .\\
0 & . & . & . & 0 & R_{n-L}^{-1}\end{array}\right]
\end{equation}

In this section we arrived at a form of filter that uses the minimum variance 
estimation method, initiated by Gauss in "Theoria Combinationis Observationum Erroribus Minimis Obnoxiae," and the local linearisation technique championed 
by Newton to estimate the state of the process from the non linear observation scheme.
This filter is called Gauss-Newton filter (GNF) and is described in detail in Morrison's work \cite{ Morrison:1969}.

\section{Adaptation to Levenberg Marquard\label{sec:adaption}}

This section represents the key step in the development of the Morrison LMA Filter. For simplicity in adaptation of the GNF to the Levenberg Marquard method we assume the dynamic of of the process we want to track is governed by linear differential equations and the observation scheme is non linear. The process transition equation will be:

\begin{equation}
X_{n+\varsigma}=\Phi(\varsigma)X_{n}\end{equation}.

The GNF will fail to converge if the matrix  $(\mathbf{T}_{n}^{T}\mathbf{R}_{n}^{-1}\mathbf{T}_{n})^{-1}$ is singular. By definition this matrix is positive definite However, it can loose this property due to numerical inaccuracy or high non-linearity. To avoid the singularity, a damping factor is introduced in equation as follows:

\begin{equation}
\delta\hat{X}_{n}=(\mathbf{T}_{n}^{T}\mathbf{R}_{n}^{-1}\mathbf{T}_{n}+\mu I)^{-1}
\mathbf{T}_{n}^{T}\mathbf{R}_{n}^{-1}\mathbf{\delta Y}_{n}
\end{equation}
which is the form suggested by Levenberg and Marquardt \cite{
     Mquardt} 

The effect of the damping factor is as follows:

\begin{itemize}
\item For all positive $\mu$ the matrix $(\mathbf{T}_{n}^{T}\mathbf{R}_{n}^{-1}\mathbf{T}_{n}+\mu I)$ is positive definite, ensuring that $\delta X$ is in the descent direction;
\item When $\mu$ is large we have:
\end{itemize}

\begin{equation}
\delta\hat{X}=\frac{1}{\mu}\mathbf{T}_{n}^{T}\mathbf{R}_{n}^{-1}\delta\mathbf{Y}_{n}
\end{equation}

The algorithm behaves as a steepest descent which is ideal when the current solution is far from the local minimum. The convergence will be slow but however guaranteed. When $\mu$ is small, the algorithm  has faster convergence and behaves like the Gauss-Newton.

The damping factor can be updated by the gain ratio:
\begin{equation}
\varrho=\frac{\mathbf{\delta Y}_{n}^{T}\mathbf{R}_{n}^{-1}\mathbf{\delta Y}_{n}-(\mathbf{Y}_{n}-\bar{\mathbf{Y}})^{T}\mathbf{R}_{n}^{-1}(\mathbf{Y}_{n}-\bar{\mathbf{Y}})}{\mathbf{E}(\delta X_{n})}
\end{equation}

where $\mathbf{\delta Y}_{n}$ is the long vector of L sequences of obserservation including the current observation.

$\bar{\mathbf{Y}}$ is the long error free observation computed by back propagation of the current iterate $X_{new}$. If  we sample at constant rate $\varsigma$ then: 

\begin{equation}
\bar{\mathbf{Y}}=\left[\begin{array}{c}
G(X_{new})\\
G(\Phi(-\varsigma)X_{new})\\
\vdots\\
G(\Phi(-(L-1)\varsigma)X_{new})\end{array}\right]
\end{equation}

The numerator is the actual computed gain and the denominator is the predicted gain. Recalling equation \ref{eq:cost_func} and replacing $\delta X_{n}$ by the expression in equation \ref{eq:deltax} we have:

\begin{equation}
E(\delta X_{n})=\delta X_{n}^{T}(\mathbf{T}_{n}^{T}\mathbf{R}_{n}^{-1}\mathbf{T}_{n})(\mathbf{T}_{n}^{T}\mathbf{R}_{n}^{-1}\mathbf{T}_{n}+\mu I)^{-1}\mathbf{T}_{n}^{T}\mathbf{R}_{n}^{-1}\delta\mathbf{Y}_{n}
\end{equation}

which reduces to:

\begin{equation}
E(\delta X_{n})=\delta X_{n}^{T}(\mathbf{T}_{n}^{T}\mathbf{R}_{n}^{-1}\delta\mathbf{Y}_{n}+\mu\delta X_{n})\end{equation}

A large value of $\varrho$ indicates that $E(\delta X_{n})$ is a good approximation of $\bar{\mathbf{Y}}$, and $\mu$ can be decreased  so that the next Levenberg-Marquardt step is closer to the Gauss-Newton step. If $\varrho$ is small or negative then $E(\delta X_{n})$ is a poor approximation, then $\mu$ should be increased to move closer to the steepest descent direction. The algorithm adapted from \cite{ LMA} is presented as follows:

\begin{algorithm}
$k:=0$,$\nu:=2$,$X:=X_{n-1/n}$

$A:=\mathbf{T}_{n}^{T}\mathbf{R}_{n}^{-1}\mathbf{T}_{n}$;$\delta\mathbf{Y}_{n}:=\mathbf{Y}_{n}-\mathbf{\bar{Y}}_{n}$;
$g:=\mathbf{T}_{n}^{T}\mathbf{R}_{n}^{-1}\delta\mathbf{Y}_{n}$;
$\mathbf{\bar{Y}}_{n}$ is computed using $X$

$stop:=false$;$\mu=\tau*max(diag(A))$;

While (not stop) and ($k\leq k_{max}$)

~~~~$k:=k+1$

~~~~repeat

~~~~solve $(A+\mu I)\delta\hat{X}_{n}=g$

~~~~if ($||\delta\hat{X}_{n}||\leq\varepsilon||X||$)

~~~~~~~~~stop:=true;

~~~~else

~~~~~~~~~~$X_{new}:=X+\delta\hat{X}_{n}$;

~~~~~~~~~~~~~$\varrho=[\delta\mathbf{Y}_{n}^{T}\mathbf{R}_{n}^{-1}\delta\mathbf{Y}_{n}-(\mathbf{Y}_{n}-\mathbf{\bar{Y}})^{T}\mathbf{R}_{n}^{-1}(\mathbf{Y}_{n}-\mathbf{\bar{Y}})]/[\delta\hat{X}_{n}^{T}(g+\mu\delta\hat{X}_{n})]$;
$\mathbf{\bar{Y}}$  evaluated at $X_{new}$

~~~~~~~~~~~~~if $\varrho>0$

~~~~~~~~~~~~~~~~~~~~~$X=X_{new}$;

~~~~~~~~~~~~~~~~~~~~~~$A:=\mathbf{T}_{n}^{T}\mathbf{R}_{n}^{-1}\mathbf{T}_{n}$;$\delta\mathbf{Y}_{n}:=\mathbf{Y}_{n}-\mathbf{\bar{Y}}_{n}$;
$g:=\mathbf{T}_{n}^{T}\mathbf{R}_{n}^{-1}\delta\mathbf{Y}_{n}$;
$\mathbf{\bar{Y}}_{n}$  is computed using X

~~~~~~~~~~~~~~~~~~~~~~$\mu=\mu*max(1/3,1-(2\varrho+1)^{3})$;$\nu:=2$;

~~~~~~~~~~~~~else

~~~~~~~~~~~~~~~~~~~~~~~$\mu:=\nu*\mu$;

~~~~~~~~~~~~~~~~~~~~~~~~$\nu:=2*\nu$;

~~~~~~~~~~~~endif

~~~~~~~endif

~~~~until($\varrho>0$)or(stop);

endwhile

$X_{n/n}=X$;

$X_{n/n+1}=\Phi(s)X_{n/n}$

\caption{L-M algorithm for tracking system}

\end{algorithm}subsec:local
\section{Simulations\label{sec:simulations}}
In the simulation studies we adopt multiple target dynamics:
\begin{itemize}
\item Case 1 : The target is moving at constant velocity under the process noise of constant standard deviation.
\item Case 2 : The target moving at constant velocity with process noise standard deviation varying

\end{itemize}
In all the cases the observation scheme is non linear. The observables are range $\rho$,bearing $\phi$,elevation 
$\theta$ and Doppler $f_{d}$. The observation equation is therefore defined as follows:

\begin{equation}
Y=\left[\begin{array}{c}
\sqrt{x^{2}+y^{2}+z^{2}}\\
tan^{-1}(y/x)\\
tan^{-1}(z/\sqrt{x^{2}+y^{2}})\\
K_{d}\frac{x\dot{x}+y\dot{y}+z\dot{z}}{\sqrt{x^{2}+y^{2}+z^{2}}}\end{array}\right]+v(t)
\end{equation}

where $v(t)$ is vector of random variables with covariance
 \[
R=\left[\begin{array}{cccc}
60^{2} & 0 & 0 & 0\\
0 & 0.001^{2} & 0 & 0\\
0 & 0 & 0.001^{2} & 0\\
0 & 0 & 0 & 2^{2}\end{array}\right]\]
throughout the simulations. $K_{d}=-2\pi/\lambda=-200$.
The constants $\tau=10^{-1}$, $\varepsilon=10^{-20}$, $k_{max}=200$, $\zeta=1s$ are used in all the cases.
\subsection{Case 1}

In this example, we seek to demonstrate that the filter does not diverge in the  presence of constant variance process which is unknown to its model.   :
The target state vector $X=[x,\dot{x},y,\dot{y},z,\dot{z}]^{T}$ is defined by the following transition equation:
\begin{equation}
X_{n+1}=\left[\begin{array}{cccccc}
1 & \varsigma & 0 & 0 & 0 & 0\\
0 & 1 & 0 & 0 & 0 & 0\\
0 & 0 & 1 & \varsigma & 0 & 0\\
0 & 0 & 0 & 1 & 0 & 0\\
0 & 0 & 0 & 0 & 1 & \varsigma\\
0 & 0 & 0 & 0 & 0 & 1\end{array}\right]X_{n}+\left[\begin{array}{c}
\frac{1}{2}a_{1}\varsigma^{2}\\
a_{1}\varsigma\\
\frac{1}{2}a_{2}\varsigma^{2}\\
a_{2}\varsigma\\
\frac{1}{2}a_{3}\varsigma^{2}\\
a_{3}\varsigma\end{array}\right]\end{equation}

where $a_{1}$, $a_{1}$, $a_{1}$ are independent,  Gaussian random variables, with standard deviation $\sigma=0.001$. The state vector is used in to generate measurements for the simulation. The filter, however, does not depend on the process noise, it assumes the target is moving at constant speed without process noise. 

The initial value of the state vector is $X=[800,25,1000,-25,400,14]$. Two thousand samples are generated and the process is repeated 50 times. The position root mean squared error (RMSE) after the 50 Monte Carlo runs is presented in Figure [1]. The position RMSE is computed as follows :
\begin{equation}RMSE=\sqrt{\frac{1}{N}\sum_{i=1}^{N}\left((x_{n}^{i}-\hat{x_{n}})^{2}+(y_{n}^{i}-\hat{y_{n}})^{2}+(z_{n}^{i}-\hat{z_{n}})^{2}\right)}\end{equation}
where $(x_{n}^{i},y_{n}^{i},z_{n}^{i})$ and $(\hat{x_{n}},\hat{y_{n}},\hat{z_{n}})$ true and estimated position coordinates respectively.

We see from Figure [1] that there is no divergence in position despite the presence of the process noise, which is unknown to the filter. The filter with the smallest memory exhibits the largest RMSE. The average number of iterations is presented in Fgure [2]. All the filters have about the same value of $k$, which is around 34, meaning the computation time of the algorithm is primarily dependent on the computation of the $\mathbf{T}_{n}$ matrix.Therefore if we want to reduce the computation time of the algorithm, we would choose a small memory length  (the $\mathbf{T}_{n}$ matrix will be small and hence less computation), but this would result in less accuracy in the estimates.

\subsection{Case 2}

Here we show the effect of higher variation in the process noise on the filter performance.
In this case the target dynamic model is the same as in Case 1. The standard deviation($\sigma$) of the process noise is varied. From sample 0 to 200 $\sigma=0.001$, between samples 201  and 260 $\sigma=0.05 $ and finally from sample 261 to 400 $\sigma=0.001$. The position RMSE after 200 Montecarlo runs is shown in Figure [3].  All the filters reset  to the original RMSE when the process noise standard deviation returned to the former value.  The RMSE of filter with the smallest memory length is less affected by these changes. However the  number of iterations during high disturbance is higher for the smaller memory length filter (Figure [4]).  Theses results highlight the adaptiveness of the algorithm to disturbance.They  also give a hint about the ability of the filter to track manoeuvre. This will be the subject of our next pubiblication.

\section{Conclusion\label{sec:conc}}
This paper introduced the standard Gauss Newton filter that  uses the  back propagation of the predicted state vector over a finite memory length to compute the Jacobian matrix. It then computes the current estimate of the state vector through the minimum variance estimation. The Gauss Newton was then adapted to the Levenberg and Marquard method to guarantee its convergence all the time. 

The adapted algorithm was used in simulations to track targets in Cartesian coordinates when the observations consist of range, bearing , elevation and Doppler. The results highlight the robustness of the new, Morrison LMA filter which can withstand strong random disturbance and nonlinear trajectories.
We observed from simulations studies that by adaptively changing the memory length, the filter will be able to track maneuvres. Such memory control algorithm will be a subject of our next publication.

\appendix
\section{}
\subsection {The differential equation governing $\delta X(t)$ \label{app:partA}}
Starting from :
\begin{equation}
\delta X(t)=X(t)-\bar{X}(t)\end{equation}
The differentiation rule is applied:

\begin{equation}
D\delta X(t)=F(\bar{X}(t)+\delta X(t))-F(\bar{X}(t))\end{equation}
Let $F$ be defined as follows :

\begin{equation}
F=\left[\begin{array}{c}
f_{1}\\
.\\
.\\
.\\
f_{n}\end{array}\right]\end{equation}
Equation becomes :
\begin{equation}
D\delta X(t)=\left[\begin{array}{c}
f_{1}(\bar{X}(t)+\delta X(t))\\
.\\
.\\
.\\
f_{n}(\bar{X}(t)+\delta X(t))\end{array}\right]-\left[\begin{array}{c}
f_{1}(\bar{X}(t))\\
.\\
.\\
.\\
f_{n}(\bar{X}(t))\end{array}\right]\end{equation}
The Taylor first order approximation is applied:
\setlength {\arraycolsep}{0.0em}
\begin{eqnarray}
D\delta X(t)&{}=&{}\left[\begin{array}{c}
f_{1}(\bar{X}(t))\\
.\\
.\\
.\\
f_{n}(\bar{X}(t))\end{array}\right]+\left[\begin{array}{c}
\nabla f_{1}(\bar{X}(t))^{T}\\
.\\
.\\
.\\
\nabla f_{n}(\bar{X}(t))^{T}\end{array}\right]\delta X(t) \nonumber \\
&&{-}\:\left[\begin{array}{c}
f_{1}(\bar{X}(t))\\
.\\
.\\
.\\
f_{n}(\bar{X}(t))\end{array}\right]\end{eqnarray}
\setlength {\arraycolsep}{5pt}
The following relation is obtained :
\begin{equation}
D\delta X(t)=A(\bar{X}(t))\delta X(t)\end{equation}
Where:
\begin{equation}
A(\bar{X}(t))=\left[\begin{array}{c}
\nabla f_{1}(\bar{X}(t))^{T}\\
.\\
.\\
.\\
\nabla f_{n}(\bar{X}(t))^{T}\end{array}\right]=\left.\frac{\partial F(X(t))}{\partial(X(t))}\right|_{\bar{X}(t)}\end{equation}

\subsection {The relation between  $\delta X_{n}$ and $\delta Y_{n}$}

\begin{equation}
\delta Y_{n}=G(\bar{X}_{n}+\delta X_{n})-G(\bar{X}_{n})\end{equation}

As direct consequence of  \ref{app:partA} the following relationship is obtained:

\begin{equation}
\delta Y_{n}=M(\bar{X}_{n})\delta X_{n}+v_{n}\end{equation}


\section*{Acknowledgment}

The authors would like to thank our colleague Dr Norman Morrison for his contribution in introducing us to the GNF and his tireless enthusiasm for teaching and  providing insights into  the fundamentals of Filter  Engineering. We wish him well for his soon to be published book, which provides new material on these remarkable filters.




%
\bibliographystyle{elsarticle-num}
\bibliography{Reference}
\newpage







\end{document}